\documentclass[preprint,showpacs,preprintnumbers,superscriptaddress]{revtex4-1}
\usepackage{amsmath,amssymb,amsthm}
\usepackage{graphicx}
\usepackage{hyperref}
\usepackage{upgreek}

\usepackage{indentfirst} % 确保第一章/第一节的第一段也缩进

\AtBeginDocument{
  \setlength{\parindent}{2em} % 再次强调缩进设置
  \noindent\ignorespaces % 取消可能的全局无缩进命令
}

\usepackage{multirow}
\usepackage{diagbox}
\usepackage{booktabs}
\usepackage{array}
\usepackage{natbib}
\usepackage{amsmath,amssymb,graphicx,bm,subfigure,float,siunitx,color}
\usepackage{bm,subfigure,float,siunitx,graphicx}
\usepackage[figuresright]{rotating}
\usepackage{color}
\usepackage{epstopdf}
\usepackage{mathrsfs}

\usepackage{multirow}
\usepackage{diagbox}
\usepackage{booktabs}
\usepackage{array}
\usepackage{natbib}
\usepackage{amsmath,amssymb,graphicx,bm,subfigure,float,siunitx,color}
\usepackage{bm,subfigure,float,siunitx,graphicx}
\usepackage[figuresright]{rotating}
\usepackage{color}
\usepackage{epstopdf}
\usepackage{mathrsfs}

\usepackage{textcomp}

\newcommand{\be}{\begin{equation}}
\newcommand{\ee}{\end{equation}}

\newcommand\bea{\begin{eqnarray}}
	\newcommand\eea{\end{eqnarray}}
\newcommand{\dd}{\mbox{d}}
\newcommand{\sn}{\mbox{s}}
\newcommand{\cn}{\mbox{c}}

\begin{document}

\title{Quantum interference in the Einstein-Maxwell-Scalar spacetime}
\author{Yingdong Wu \footnote{Corresponding author}}
\email{Email address: yingdongwu7@gmail.com}
%\affiliation{Department of Physics, State key laboratory of quantum functional materials, and Guangdong Basic Research Center of Excellence for Quantum Science, Southern University of Science and Technology, Shenzhen 518055, China}
\affiliation{State Key Laboratory of Quantum Functional Materials, Department of Physics, and Guangdong Basic Research Center of Excellence for Quantum Science, Southern University of Science and Technology, Shenzhen 518055, China}

\author{Wei-Qiang Chen \footnote{Corresponding author} }
\email{Email address: chenwq@sustech.edu.cn}
%\affiliation{Department of Physics, State key laboratory of quantum functional materials, and Guangdong Basic Research Center of Excellence for Quantum Science, Southern University of Science and Technology, Shenzhen 518055, China}
\affiliation{State Key Laboratory of Quantum Functional Materials, Department of Physics, and Guangdong Basic Research Center of Excellence for Quantum Science, Southern University of Science and Technology, Shenzhen 518055, China}

\affiliation{Quantum Science Center of Guangdong-Hong Kong-Macao Greater Bay Area, Shenzhen 518045, China}

\begin{abstract}
We investigate gravitationally induced interference within the framework of 
Teleparallel Gravity, deriving a general expression for the gravitational phase 
difference and applying it to an Einstein–Maxwell–Scalar (EMS) spacetime. We 
then analyze how this phase difference is affected by the presence or absence 
of black hole charge. Our results show that, irrespective of charge, the 
dominant contribution to the gravitational phase difference arises from the 
black hole mass. Nevertheless, while the influence of charge is negligible 
under standard astrophysical conditions, we identify a possible amplification 
mechanism through its interplay with the coupling parameter $\alpha$ in the 
case of extremal charged black holes. This finding not only offers new 
theoretical insights but also suggests experimental opportunities for probing 
charged black hole parameters via quantum interferometric techniques.

\end{abstract}
\maketitle

\section{Introduction\label{Intro}}
The Aharonov--Bohm (AB) effect, proposed in 1959 by Aharonov and 
Bohm~\cite{Aharonov:1959fk}, demonstrates that electromagnetic potentials can 
influence quantum systems even in regions where the corresponding fields 
vanish. In their first thought experiment, two electron beams travel through 
separate cylindrical tubes with different time-dependent scalar potentials and 
then interfere in a field-free region. In a second proposal, an electron beam 
is split into two paths encircling an infinite solenoid that confines a 
magnetic field to its interior. In both cases, the relative phase shift is 
determined by the potential, not the local field.
%and can be written as
%\begin{equation}
%	-\frac{1}{\hbar}(\mathscr{S}_1-\mathscr{S}_2)
%	= \frac{e}{\hbar c} \oint A_\mu\, dx^\mu ,
%	\label{eq:AB-phase}
%\end{equation}
%where $A_\mu$ is the electromagnetic four-potential. 
The magnetic AB effect was first observed by Chambers~\cite{Chambers:1960xlk} and later confirmed with high precision~\cite{Tonomura:1989zz}.

A gravitational analogue of the AB effect has been 
proposed~\cite{Dowker,Ford,Audretsch:1982ux}. In this case, a quantum particle 
moves through a region of spacetime where the Riemann curvature tensor vanishes 
locally, but the global spacetime structure induces a phase shift. In the 
semiclassical limit, Stodolsky~\cite{Stodolsky:1978ks} showed that the 
gravity-induced phase for a massive particle along a classical trajectory is
\begin{equation}
	\frac{mc}{\hbar} \int ds ,
	\label{eq:Stodolsky-phase}
\end{equation}
which is invariant under general coordinate transformations, in contrast to the 
$U(1)$ gauge dependence of the electromagnetic AB phase.

Teleparallel gravity (TG), or the teleparallel equivalent of general relativity 
(TEGR)~\cite{Aldrovandi:2013wha}, provides a natural framework to express the 
gravitational phase in gauge-theoretic form. In TG, gravitation is described by 
torsion rather than curvature, with the tetrad field as the fundamental 
variable. The phase in Eq.~\eqref{eq:Stodolsky-phase} can be decomposed into 
three parts~\cite{Aldrovandi:2013wha}: a free-particle term, a purely inertial 
term removable by an appropriate frame choice, and a genuine gravitational term.
%\begin{equation}
%	\frac{mc}{\hbar} \int u_a {B^a}_\mu\, dx^\mu ,
%	\label{eq:TG-phase}
%\end{equation}
%where ${B^a}_\mu$ is the translational gauge potential and $u_a$ is the tetrad-projected four-velocity. 
In the weak-field limit, this expression 
reproduces the gravitational phase observed in the Colella--Overhauser--Werner 
(COW) neutron interferometry experiment~\cite{Colella:1975dq}.

Gravitationally induced quantum interference has been measured in a variety of 
laboratory settings. The COW experiment demonstrated a phase shift proportional 
to Earth’s gravitational acceleration~\cite{Colella:1975dq}. More recently, 
Overstreet \emph{et al.}~\cite{Overstreet:2021hea} used atom interferometry to 
detect the phase shift produced by small laboratory-scale masses, marking a 
significant step toward precision measurements of gravitational phases.  In addition, the recent paper \cite{207} has, for the first time, provided high-precision verifications of two fundamental theoretical predictions of black holes: Hawking’s area theorem and the “no-hair” (Kerr) property. Paper \cite{203} also specifically discusses the influence of the Kerr property in quantum interference.

While most experiments focus on weak gravitational fields, strong-field 
environments such as black holes provide a natural arena to test gravitational 
phase effects. Although astrophysical black holes are usually expected to be 
nearly neutral due to rapid charge neutralization by surrounding plasma, recent 
estimates~\cite{202} suggest that newly formed stellar-mass black holes could 
possess charges as large as $\mathcal{O}(10^{13})$~C. This motivates the study 
of gravitational phases in charged black hole spacetimes. Namely, the third possible hair of a black hole—charge—and the effects it can produce.  In particular, we will focus on the 
Einstein--Maxwell--Scalar (EMS) model which generalizes the Reissner--Nordström 
geometry and allows us to investigate how the black hole charge and 
scalar--electromagnetic coupling parameters affect the gravitational phase.

The Einstein–Maxwell–Scalar (EMS)   model \cite{208,Feng_2022} extends the standard Einstein–Maxwell theory by introducing a nonminimal coupling between the scalar and electromagnetic fields. This coupling modifies the electromagnetic energy–momentum tensor and, consequently, the spacetime geometry. In the static, spherically symmetric sector, the EMS solution reduces to the Reissner–Nordström (RN) black hole when the coupling parameter $\alpha, \beta$ vanishes. In particular, the effective metric components acquire additional $\alpha,\beta$-dependent terms that can enhance or suppress the contribution of the black hole charge $Q$ to observables. This leads to the intriguing possibility that, under certain conditions, the combined $Q$–$\alpha$-$\beta$ term may dominate over the mass term in physical effects—something impossible in the RN spacetime within general relativity.

%The teleparallel gravity (TG) formalism offers a particularly convenient approach for studying such scenarios. Unlike the curvature-based formulation of general relativity, TG encodes gravitation in torsion, described by the tetrad field and the translational gauge potential ${B^a}_\mu$. This makes the gravitational phase shift naturally expressible in gauge-theoretic terms, analogous to the electromagnetic AB effect, and allows a clean separation between inertial and gravitational contributions. This feature is essential when assessing how modifications to the spacetime geometry—such as those induced by the EMS coupling—translate into measurable phase shifts.

%Previous studies of gravitational Aharonov–Bohm–type effects have largely concentrated on either: (i) weak-field laboratory experiments, where the gravitational field is nearly Newtonian, or (ii) standard GR black hole metrics, most often the Schwarzschild or RN spacetimes. The role of additional fields and couplings—particularly scalar–electromagnetic interactions—remains poorly explored in the context of gravitational phase shifts. As a result, there is currently no quantitative understanding of how strong-field modifications to the geometry impact the interference patterns predicted by TG-based phase formulas.

In this work, we aim to answer the following scientific question:
How does the presence of a nonminimally coupled scalar field in the EMS model modify the gravitational phase shift of a test particle, and under what conditions can the $Q$–$\alpha$-$\beta$ term become the dominant contribution?
We provide a quantitative analysis by deriving an explicit inertial-frame expression for the gravitational phase in the EMS spacetime, evaluating its dependence on $(M, Q, \alpha,\beta)$, and identifying the regions of parameter space where
\begin{equation}
    \frac{\Delta\phi_{Q,\alpha,\beta}}{\Delta\phi_M} \gtrsim 1 ,
\end{equation}
indicating that the charge–coupling term equals or exceeds the mass term in magnitude. This not only generalizes existing TG phase calculations to a broader class of black hole solutions but also offers testable predictions for future strong-field gravitational interference experiments.

In this work, we use the TG formalism to compute the gravitational phase for 
test particles in the EMS spacetime. We derive an explicit inertial-frame 
expression for the phase, evaluate its dependence on the black hole parameters, 
and discuss possible observational implications in a gravitational 
Aharonov--Bohm--type interference experiment. The paper is organized as 
follows: Sec.~II reviews the tetrad formalism in TG; Sec.~III presents the 
derivation of the gravitational phase in inertial frames and applies it to the 
EMS solution; Sec.~IV analyzes the resulting phase shift in an interference 
setup; and Sec.~V summarizes our findings.

Throughout this article, we use the units $c=G=1$ and the metric signature 
$(+,-,-,-)$, unless we explicitly specify.

%%%%%%%%%%%%%%%%%%%%%%%%%%%%%%%%%%%%%%%%%%%%%
%%%%%%%%%%%%%%%%%%%%%%%%%%%%%%%%%%%%%%%%%%%%%%%

\section{A brief introduction to Teleparallel Gravity and EMS theory}
In this chapter, to aid the derivation and readability of the paper, we will follow the relevant section of reference \cite{203} and give a brief introduction to teleparallel gravity and EMS theory. 

\subsection{Tetrad formulation and the dynamical role of the translational gauge potential}
We summarize the key elements of TG following Ref.~\cite{Aldrovandi:2013wha}, focusing on the tetrad formalism and the dynamical role of the translational gauge potential ${B^a}_\mu$.

The tetrad fields ${h_a}^\mu$ and their duals ${h^a}_\mu$ define a local orthonormal frame at each spacetime point, relating the spacetime metric $g_{\mu\nu}$ to the Minkowski metric $\eta_{ab}$:
\begin{equation}
g_{\mu\nu} = \eta_{ab} {h^a}_\mu {h^b}_\nu,
\quad
\eta_{ab} = g_{\mu\nu} {h_a}^\mu {h_b}^\nu,
\label{metric transformation}
\end{equation}
with orthonormality conditions:
\begin{equation}
{h^a}_\mu {h_a}^\nu = \delta^\nu_\mu,
\quad
{h^a}_\mu {h_b}^\mu = \delta^a_b.
\label{tetrad components}
\end{equation}
Tensor quantities can be mapped between spacetime and tangent space via:
\begin{equation}
V^\mu = {h_a}^\mu V^a,
\quad
V_a = {h^a}_\mu V^\mu.
\label{tensor transformation}
\end{equation}

In TG, the tetrad is decomposed as\cite{Aldrovandi:2013wha}:
\begin{equation}
{h^a}_\mu = \partial_\mu x^a + \dot{A}{^a}_{b\mu} x^b + {B^a}_\mu,
\label{tetrad}
\end{equation}
where $\dot{A}{^a}_{b\mu} = {\Lambda^a}_d \partial_\mu {\Lambda_b}^d$ is the inertial Lorentz connection and ${B^a}_\mu$ is the translational gauge potential encoding gravitational effects. Under local translations $\delta x^a = \varepsilon^a(x)$, the gauge potential transforms as:
\begin{equation}
\delta {B^a}_\mu = -\partial_\mu \varepsilon^a - \dot{A}{^a}_{b\mu} \varepsilon^b.
\label{gauge B}
\end{equation}
In differential form language, the tetrad 1-form can be expressed as
\begin{equation}
	h^a = \mathrm{d} x^a + \dot{A}{^a}_{b\mu} x^b \mathrm{d} x^\mu + B{^a}_\mu \mathrm{d} x^\mu.
	\label{ha expression}
\end{equation}

TG employs the Weitzenböck connection, which is curvature-free but has non-vanishing torsion:
\begin{align}
\dot{R}^a{}_{b\mu\nu} &= 0,
\label{dotR}
\\
\dot{T}^a{}_{\mu\nu} &= \dot{\mathscr{D}}_\mu {B^a}_\nu - \dot{\mathscr{D}}_\nu {B^a}_\mu,
\label{dotT}
\end{align}
where $\dot{\mathscr{D}}_\mu \phi^a = \partial_\mu \phi^a + \dot{A}{^a}_{b\mu} \phi^b$. The torsion tensor is invariant under the gauge transformation above and plays the role of the gravitational field strength.

The motion of a test particle in TG is governed by an equation equivalent to the geodesic equation of GR. The Levi–Civita connection is decomposed as\cite{Aldrovandi:2013wha}:
\begin{equation}
\Gamma^\mu{}_{\rho\nu} = \dot{\Gamma}^\mu{}_{\rho\nu} - \dot{K}^\mu{}_{\rho\nu},
\label{GammaK}
\end{equation}
with the contortion tensor:
\begin{equation}
\dot{K}^\mu{}_{\rho\nu} = \frac{1}{2} \left( \dot{T}_\nu{}^\mu{}_\rho + \dot{T}_\rho{}^\mu{}_\nu - \dot{T}^\mu{}_{\rho\nu} \right),
\label{Kdefiniton}
\end{equation}
constructed from the torsion. Substituting into the geodesic equation yields:
\begin{equation}
\frac{d u^\mu}{ds} + \left( \dot{\Gamma}^\mu{}_{\rho\nu} - \dot{K}^\mu{}_{\rho\nu} \right) u^\rho u^\nu = 0,
\label{EOMu}
\end{equation}
which, in terms of the tetrad and gauge potential, becomes:
\begin{equation}
\frac{d u^\mu}{ds} + {h_a}^\mu \left[ \dot{\mathscr{D}}_\nu {h^a}_\rho - \left( \eta_{ce} {h^c}_\rho \right) \left( \eta^{af} {h_f}^\beta \right) \left( \dot{\mathscr{D}}_\beta {B^e}_\nu - \dot{\mathscr{D}}_\nu {B^e}_\beta \right) \right] u^\rho u^\nu = 0.
\label{duds}
\end{equation}
This confirms that ${B^a}_\mu$ encodes the gravitational interaction and cannot be gauged away.

The gauge potential also enters the particle action\cite{Aldrovandi:2013wha}:
\begin{equation}
\mathscr{S} = - m \int \left( u_a dx^a + u_a \dot{A}{^a}_{b\mu} x^b dx^\mu + u_a {B^a}_\mu dx^\mu \right),
\label{S action}
\end{equation}
where the last term represents the gravitational interaction. In analogy with the electromagnetic Aharonov–Bohm phase, the gravitational phase factor in TG is determined by the integral of ${B^a}_\mu$, highlighting its role as the true gravitational potential.

%%%%%%%%%%%%%%%%%%%%%%%%%%%%%%%%%%%%%%%%%%%%%%%%%%%%%%%%%%%%%%%%%%%%%%%%%%%%%%%%%
%%%%%%%%%%%%%%%%%%%%%%%%%%%%%%%%%%%%%%%%%%%%%%%%%%%%%%

\subsection{Einstein-Maxwell-Scalar Theory}
The systematic investigation of scalar-gravity coupling was introduced by 
Fisher, who obtained static spherically symmetric solutions to Einstein's 
equations with a massless scalar field \cite{102}. This foundational work 
established the theoretical basis for subsequent developments, particularly in 
Einstein-Maxwell-Scalar (EMS) theory which has attracted significant attention 
due to its implications and natural emergence in: (i) Kaluza-Klein 
dimensional reduction \cite{103}, (ii) dilaton couplings in string theory 
\cite{104}, and (iii) cosmological inflation and late-time acceleration 
\cite{105}.

The EMS action takes the form
\begin{equation}
	\label{1}
	S = \frac{1}{16\pi}\int d^4x\sqrt{-g}\left[R - 2(\nabla\phi)^2 - K(\phi)F^2 
	- V(\phi)\right],
\end{equation}
where $R$ is the Ricci scalar, $\phi$ is the scalar field. 
$F_{\mu\nu}=\partial_{\mu}A_{\nu}-\partial_{\nu}A_{\mu}$ represents the 
electro-magnetic field tensor. The field equations derived from this action 
yield
\begin{equation}
	\begin{cases}
		\nabla_\mu[K(\phi)F^{\mu\nu}] = 0, \\
		\square \phi = \frac{1}{4}\left[\partial_\phi K(\phi)F^2+\partial_\phi 
		V(\phi) \right], \\
		R_{\mu\nu} = 2\partial_\mu\phi\partial_\nu\phi + 
		\frac{1}{2}g_{\mu\nu}V(\phi) +2K(\phi)(F_{\mu\sigma}F^\sigma_{~\nu} - 
		\frac{1}{4}g_{\mu\nu}F^2).
	\end{cases}
\end{equation}

The coupling function $K(\phi)$ and potential $V(\phi)$ determine the theory's 
physical content. Some important solutions with different $K(\phi)$ and 
$V(\phi)$ are given by \cite{3,4,5,6,7,8}. Notable special solutions include
\begin{itemize}
	\item RN-de Sitter BH :$K(\phi)=1$, $V=2\Lambda$,
	\item Dilaton BH :$K(\phi)=e^{2\phi}$, $V=0$ \cite{1,2}.
\end{itemize}

Our investigation focuses on the parameterized coupling function
\begin{equation}
	\label{5}
	K(\phi) = 
	\frac{(\alpha^2+1)e^{-2\phi/\alpha}}{(\alpha^2+1+\beta)e^{-2\phi(\alpha^2+1)/\alpha}
	 + \beta\alpha^2},
\end{equation}
which exhibits two physically significant limits
\begin{itemize}
	\item $\beta\to\infty$: Decoupled electromagnetic sector
	\item $\beta\to 0$: Einstein-Maxwell-Dilaton (EMD) theory with 
	$K(\phi)=e^{2\alpha\phi}$
\end{itemize}

As shown in Ref.\cite{9}, these admit BH solution with metric components
\begin{equation}
	\begin{cases}
		ds^2 = f(r)dt^2 -f^{-1}(r)dr^2 - C(r)d\Omega^2, \\
		f(r) = 
		\left(1-\frac{b_1}{r}\right)\left(1-\frac{b_2}{r}\right)^{\frac{1-\alpha^2}{1+\alpha^2}}
		 + \frac{\beta Q^2}{C(r)}, \\
		C(r) = r^2\left(1-\frac{b_2}{r}\right)^{\frac{2\alpha^2}{1+\alpha^2}},
	\end{cases}
\end{equation}
where the parameters $b_1$ and $b_2$ are determined by
\begin{equation}
	\begin{cases}
		b_1 = \left(1 + \sqrt{1 - q^2(1-\alpha^2)}\right)M ,\\
		b_2 = \frac{1+\alpha^2}{1-\alpha^2}\left[1 - \sqrt{1 - 
		q^2(1-\alpha^2)}\right]M,
	\end{cases}
\end{equation}
with $q\equiv Q/M$ representing the dimensionless charge-to-mass ratio.

%%%%%%%%%%%%%%%%%%%%%%%%%%%%%%%%%%%%%%%%%%%%%%%5
%%%%%%%%%%%%%%%%%%%%%%%%%%%%%%%%%%%%%%%%%%%%%%%%%%%
\section{Gravitational phase\label{GPF}}
In this section we will briefly introduce the general formula for the gravitational 
phase following Ref.\cite{203}  and afterwards we evaluate it explicitly for the case of the EMS spacetime. 
\subsection{Gravitational phase in inertial references}

The gravitational phase factor for a massive particle in a generic frame is~\cite{Aldrovandi:2013wha,Aldrovandi:2003pd,203}:
\begin{equation}
\Phi_g = \exp \left(-\frac{i}{\hbar} \mathscr{S}_g \right),
\label{Phi g}
\end{equation}
where the interaction action is
\begin{equation}
\mathscr{S}_g = -m \int_p^q u_a \left(\dot{A}{^a}_{b\mu} x^b \dd x^\mu + B{^a}_\mu \dd x^\mu \right),
\label{Sg}
\end{equation}
and the total action reads
\begin{equation}
\mathscr{S} = -m \int_p^q \dd s.
\label{action int}
\end{equation}
The four-velocities in spacetime and tangent space are defined as
\begin{equation}
u^\mu = \frac{\dd x^\mu}{\dd s}, \quad u^a = \frac{\dd x^a}{\dd s}.
\label{four velocities}
\end{equation}

Choosing an inertial coordinate system $K$ such that $\dot{A}{^a}_{b\mu} = 0$, Eq.~\eqref{Sg} reduces to
\begin{equation}
\mathscr{S}_g = -m \int_p^q u_a B{^a}_\beta \dd x^\beta = -m \int_p^q g_{\mu\nu} u^\mu B{^\nu}_\beta \dd x^\beta,
\label{SgKK}
\end{equation}
where $B{^\nu}_\beta$ is defined as
\begin{equation}
B{^\nu}_\beta = h_a{^\nu} B{^a}_\beta = (h^T B){^\nu}_\beta,
\label{B spacetime}
\end{equation}
with $h^T$ the transpose of $h_a{^\nu}$ and $B$ the matrix with components $B{^a}_\beta$.  
Substituting \eqref{SgKK} into \eqref{Phi g} yields the gravitational phase factor for a massive particle,
\begin{equation}
\Phi_g = \exp\left(\frac{i}{\hbar} m \int_p^q g_{\mu\nu} u^\mu B{^\nu}_\beta \dd x^\beta \right).
\label{Phi mass}
\end{equation}

For massless particles, consider first photons whose phase factor is
\begin{equation}
\Phi = \exp(i \psi) = \exp\left(\frac{i}{\hbar} \int_p^q P_\mu \dd x^\mu \right),
\label{Phi L}
\end{equation}
where $P_\mu = \hbar k_\mu$ is the four-momentum and $k_\mu$ the wave vector.  
Extracting the gravitational part in the TG framework without a weak-field approximation, from \eqref{Phi L} we have
\begin{equation}
\dd \psi = k_\mu \dd x^\mu = k_a h^a = k_a \left(\dd x^a + \dot{A}{^a}_{b\mu} x^b \dd x^\mu + B{^a}_\mu \dd x^\mu \right),
\label{d psi}
\end{equation}
where relations \eqref{tensor transformation}–\eqref{ha expression} apply.  
Focusing on the interaction term and choosing $\dot{A}{^a}_{b\mu} = 0$, the gravitational phase reduces to
\begin{equation}
\phi_g = \int_p^q k_a B{^a}_\mu \dd x^\mu = \int_p^q k_\nu B{^\nu}_\mu \dd x^\mu,
\label{psi g}
\end{equation}
which leads to the gravitational phase factor for light,
\begin{equation}
\Phi_L = \exp(i \phi_g) = \exp\left(\frac{i}{\hbar} \int_p^q g_{\mu\nu} P^\mu B{^\nu}_\beta \dd x^\beta \right).
\label{Phi light}
\end{equation}
This expression is assumed valid for other massless particles, though a rigorous proof is needed.

In summary, the gravitational phase for any particle in an inertial frame is
\begin{equation}
\phi_g = \frac{1}{\hbar} \int_p^q S_\beta \dd x^\beta,
\label{phi_g}
\end{equation}
with
\begin{equation}
S_\beta = g_{\mu\nu} P^\mu B{^\nu}_\beta,
\label{defining S}
\end{equation}
and four-momentum $P^\mu$. The phase factor is $\Phi_g = \exp(i \phi_g)$.

To evaluate $S_\beta$, we first find $B{^\nu}_\beta$ from $h_a{^\nu}$ and $B{^a}_\beta$ as in \eqref{B spacetime}.  
In the Cartesian coordinate system $K'$ where $\partial_{\mu'} x^a = \delta{^a}_{\mu'}$~\cite{Aldrovandi:2013wha}, the gauge potential is
\begin{equation}
B{^a}_{\mu'} = h{^a}_{\mu'} - \delta{^a}_{\mu'}.
\label{fieldK}
\end{equation}
Transforming to a general coordinate system $K$ via
\begin{equation}
h{^a}_\rho = h{^a}_{\nu'} \frac{\partial x^{\nu'}}{\partial x^\rho},
\label{hhprime}
\end{equation}
one obtains
\begin{equation}
B{^a}_\beta = h{^a}_\beta - \delta{^a}_{\mu'} \frac{\partial x^{\mu'}}{\partial x^\beta}.
\label{B field}
\end{equation}

Thus, the computational procedure is:
\begin{enumerate}
    \item Choose an inertial frame $K$.
    \item Determine the tetrad components $h^a{_\beta}$ and coordinate transformation from $K$ to Cartesian frame $K'$.
    \item Compute $B{^a}_\beta$ via \eqref{B field}.
    \item Calculate $B{^\nu}_\beta$ using \eqref{B spacetime}.
    \item Obtain $S_\beta$ from \eqref{defining S} and perform the integral in \eqref{phi_g}.
\end{enumerate}
This framework allows a unified and exact calculation of the gravitational phase factor for both massive and massless particles within teleparallel gravity. 

%%%%%%%%%%%%%%%%%%%%%%%%%%%%%%%%%%%%%%%%%%%%%%%%%%%%
%%%%%%%%%%%%%%%%%%%%%%%%%%%%%%%%%%%%%%%%%%%%%%%%%%%%%%

\subsection{Gravitational phase in the EMS spacetime}
In the spherical, static and isotropic coordinate system
$X^{1}= \rho \sin\theta \cos\phi$, $X^{2}= \rho \sin\theta \sin\phi$,
$X^{3}= \rho \cos\theta$, the tetrad components of the EMS spacetime
can be obtained from the line element
\be
ds^{2}\equiv g_{\mu\nu}dX^{\mu}dX^{\nu}
= D(\rho)dt^{2}- E(\rho)(d\rho^{2} + \rho^{2}d\Omega^{2}) ,
\label{56}
\ee
where
\be
d\Omega^{2}= d\theta^{2} + \sin^{2}\theta d\phi^{2} .
\ee

With the subscript $\mu$ denoting the column index, they are given by 
\cite{200,201}
\begin{equation}
	h^a{ }_\mu \equiv\left(\begin{array}{cccc}
		\sqrt{D} & 0 & 0 & 0 \\
		0 & \sqrt{E} & 0 & 0 \\
		0 & 0 & \sqrt{E} & 0 \\
		0 & 0 & 0 & \sqrt{E}
	\end{array}\right),
\end{equation}
with the inverse
\begin{equation}
	h_a{}^\mu \equiv\left(\begin{array}{cccc}
		\sqrt{D^{-1}} & 0 & 0 & 0 \\
		0 & \sqrt{E^{-1}} & 0 & 0 \\
		0 & 0 & \sqrt{E^{-1}} & 0 \\
		0 & 0 & 0 & \sqrt{E^{-1}}
	\end{array}\right) .
\end{equation}

Now, the EMS  geometry can also be globally represented by the  coordinate 
system $\left\{X^\mu\right\}=$ $(t, r, \theta, \phi)$, with the line element in 
this case given by
\begin{equation}
	ds^2 = f(r)dt^2 - f^{-1}(r)dr^2 - C(r)d\Omega^2. \label{60}
\end{equation}
Comparing the line elements in the isotropic and in the EMS
coordinates, given respectively by Eqs.(\ref{56}) and (\ref{60}), we see that 
\begin{equation}
	D(\rho)=g_{00}, \quad \sqrt{E(\rho)} \rho=\sqrt{C(r)}, \quad \frac{\partial 
	\rho}{\partial r}=\sqrt{\frac{-g_{11}}{E(\rho)}} .
\end{equation}
Using the general coordinate transformation
\begin{equation}
	h^a{ }_\mu=\frac{\partial X^{\nu^{\prime}}}{\partial X^\mu} h^a{ 
	}_{\nu^{\prime}},
\end{equation}
\begin{equation}
	\frac{\partial X^{\nu^{\prime}}}{\partial X^\mu}=\left(\begin{array}{cccc}
		1 & 0 & 0 & 0 \\
		0 & \frac{\partial \rho}{\partial r} \sn \theta \cn \phi & \rho \cn 
		\theta \cn \phi & -\rho \sn \theta \sn \phi \\
		0 & \frac{\partial \rho}{\partial r} \sn \theta \sn \phi & \rho \cn 
		\theta \sn \phi & \rho \sn \theta \cn \phi \\
		0 & \frac{\partial \rho}{\partial r} \cn \theta & -\rho \sn \theta & 0
	\end{array}\right),
\end{equation}
where $\left\{X^\mu\right\}$ and $\left\{X^{\nu^{\prime}}\right\}$ are 
respectively the isotropic and EMS coordinates. We obtain the tetrad in the EMS 
coordinate system:
\begin{equation}
	h^a{ }_\mu \equiv\left(\begin{array}{cccc}
		\gamma_{00} & 0 & 0 & 0 \\
		0 & \gamma_{11} \mathrm{~s} \theta \mathrm{c} \phi & \gamma_{22} 
		\mathrm{c} \theta \mathrm{c} \phi & -\gamma_{22} \mathrm{~s} \theta 
		\mathrm{~s} \phi \\
		0 & \gamma_{11} \mathrm{~s} \theta \mathrm{~s} \phi & 
		\gamma_{22}\mathrm{c} \theta \mathrm{~s} \phi & \gamma_{22}\mathrm{~s} 
		\theta \mathrm{c} \phi \\
		0 & \gamma_{11} \mathrm{c} \theta & -\gamma_{22} \mathrm{~s} \theta & 0
	\end{array}\right),\label{64}
\end{equation}
where we have introduced the following notations: $\gamma_{00}=\sqrt{g_{00}}, 
\gamma_{i i}=\sqrt{-g_{i i}}, \mathrm{~s} \theta=\sin \theta$, and $\mathrm{c} 
\theta=\cos \theta$. Its inverse is
\begin{equation}
	h_a{ }^\mu \equiv\left(\begin{array}{cccc}
		\gamma_{00}^{-1} & 0 & 0 & 0 \\
		0 & \gamma_{11}^{-1} \mathrm{~s} \theta \mathrm{c} \phi & 
		\gamma_{22}^{-1} \mathrm{c} \theta \mathrm{c} \phi & -(\gamma_{22} 
		\mathrm{~s} \theta)^{-1} \mathrm{~s} \phi \\
		0 & \gamma_{11}^{-1} \mathrm{~s} \theta \mathrm{~s} \phi & 
		\gamma_{22}^{-1} \mathrm{c} \theta \mathrm{~s} \phi & (\gamma_{22} 
		\mathrm{~s} \theta)^{-1} \mathrm{c} \phi \\
		0 & \gamma_{11}^{-1} \mathrm{c} \theta & -\gamma_{22}^{-1} \mathrm{~s} 
		\theta & 0
	\end{array}\right) .\label{65}
\end{equation}

We will calculate the gravitational phase in the EMS spacetime by using the 
last expression in \eqref{SgKK}, but before that, we derive the expression of 
$B{^a}_\beta$ according to Eq.~\eqref{B field}. 
The coordinate transformation from $K'$ to $K$ is~\cite{Landau:1975pou}:
\be
\begin{cases}
	t'=t,\\
	x'=r\, \sn\theta\cn\phi,\\
	y'=r\, \sn\theta\sn\phi,\\
	z'=r\, \cn\theta.
\end{cases}
\label{transformation}
\ee
From Eq.~\eqref{transformation} we can get the Jacobi matrix:
\be
\Bigl(\frac{\partial x^{\mu'}}{\partial x^\beta}\Bigr)=
\begin{pmatrix}
	1 & 0 & 0 &0\\
	0& r\sn\theta \cn\phi &r \cn\theta\cn \phi &-r\sn\theta\sn\phi\\
	0& r\sn\theta \sn\phi &r \cn\theta\sn \phi &r\sn\theta\cn\phi\\
	0&\cn\theta &-r\sn\theta &0
\end{pmatrix}.
\label{Jacobi}
\ee

Inserting Eqs. \eqref{64} and \eqref{Jacobi} into \eqref{B field}, we get the 
gauge potential in the EMS spacetime:
\be
(B{^a}_\beta)=
\begin{pmatrix}
	\gamma_{00}-1 & 0 & 0 &0\\
	0 & (\gamma_{11}-r) \mathrm{~s} \theta \mathrm{c} \phi & (\gamma_{22}-r) 
	\mathrm{c} \theta \mathrm{c} \phi & (-\gamma_{22}+r )\mathrm{~s} \theta 
	\mathrm{~s} \phi \\
	0 & (\gamma_{11}-r) \mathrm{~s} \theta \mathrm{~s} \phi & 
	(\gamma_{22}-r)\mathrm{c} \theta \mathrm{~s} \phi & 
	(\gamma_{22}-r)\mathrm{~s} \theta \mathrm{c} \phi \\
	0 & (\gamma_{11}-1) \mathrm{c} \theta & (-\gamma_{22}+r) \mathrm{~s} \theta 
	& 0
\end{pmatrix}.
\label{B EMS}
\ee
Inserting Eqs.~\eqref{B EMS} and \eqref{65} into \eqref{B spacetime}, we obtain
\be
(B{^\nu}_\beta)=
\begin{pmatrix}
	1-\gamma_{00}^{-1} & 0 & 0 &0\\
	0& 1-\gamma_{11}^{-1}( r (\sn \theta)^{2}+(\cn \theta)^{2}) & 0 &0\\
	0&  \gamma_{22}^{-1}\sn\theta\cn\theta(1-r)& 1-\gamma_{22}^{-1}r &0\\
	0&0 &0 &1-\gamma_{22}^{-1}r
\end{pmatrix}.
\label{B expression}
\ee

In terms of Eq.~\eqref{defining S}, the matrix $S_\beta$ can be written as:
\be
(S_\beta)=(P^\mu g_{\mu\nu}B{^\nu}_\beta)=
\begin{pmatrix}
	(1-\gamma_{00}^{-1})P^0 g_{00}\\
	P^1 g_{11}(1-\gamma_{11}^{-1}( r (\sn \theta)^{2}+(\cn \theta)^{2})) +P^2 
	g_{22}(\gamma_{22}^{-1}\sn\theta\cn\theta(1-r))\\
	(1-\gamma_{22}^{-1}r)P^{2} g_{22}\\
	(1-\gamma_{22}^{-1}r)P^{3} g_{33}
\end{pmatrix}.
\label{S beta}
\ee
The latter result can be further simplified by using the following conserved 
quantities in the EMS  spacetime, 
\bea
E&=&u^0 g_{00}, \nonumber\\
L&=&u^3 g_{33},
\label{conserved quantities}
\eea
where $u^{\mu}$ is four-velocity, and $E$ and $L$ are defined as
\be
E=
\begin{cases}
	\mathcal{E} m^{-1}, &\text{for massive particles,}\\
	\mathcal{E}, &\text{for massless particles,}
\end{cases}
\qquad
L=
\begin{cases}
	\mathcal{L} m^{-1}, &\text{for massive particles,}\\
	\mathcal{L}, &\text{for massless particles.}
\end{cases}
\label{defining EL}
\ee
Notice that for massive particles we have $P^\mu=m u^\mu$, while for massless 
particles we have $P^\mu=u^\mu$. Here the quantity $\mathcal{E}$ has the 
meaning of energy, while the quantity $\mathcal{L}$ has the meaning of angular 
momentum. Plugging \eqref{conserved quantities} into \eqref{S beta}, we get
\be
(S_\beta)=(P^\mu g_{\mu\nu}B{^\nu}_\beta)=
\begin{pmatrix}
	(1-\gamma_{00}^{-1})\mathcal{E}\\
	P^1 g_{11}(1-\gamma_{11}^{-1}( r (\sn \theta)^{2}+(\cn \theta)^{2})) +P^2 
	g_{22}(\gamma_{22}^{-1}\sn\theta\cn\theta(1-r))\\
	(1-\gamma_{22}^{-1}r)P^{2} g_{22}\\
	(1-\gamma_{22}^{-1}r)\mathcal{L}
\end{pmatrix}.
\label{S beta2}
\ee
Finally, recalling \eqref{phi_g}, the gravitational phase in the EMS spacetime 
is given by:
\be
\phi_{g}=\frac{1}{\hbar}\int_p^q S_\beta \dd x^\beta.
\label{PhiK}
\ee  

\section{PARTICLES INTERFERENCE EXPERIMENT\label{interference}}
\subsection{Theoretical modeling and physical predictions\label{TPre}}
We examine a quantum-interference configuration situated far from a black hole's horizon ($r \gg r_{g}$), where the size of the apparatus is much smaller than its distance from the black hole. We use the well-known Colella-Overhauser-Werner (COW) experiment \cite{Colella:1975dq} as our terrestrial reference.
In the COW experiment, a neutron interferometer is set up as a vertical parallelogram. A neutron beam is split and travels along two paths, ABC and ADC, before recombining to create an interference pattern. Due to the uniform gravitational field, a phase difference emerges between the two paths, resulting in a measurable phase shift.
The gravitationally induced phase shift, $\delta\phi_{g}$, is given by\cite{Overhauser:1974}:
\be
\delta\phi_{g}=\frac{m^2 g l \lambda_d}{2\pi\hbar^2}s,
\label{COWpre}
\ee
here, \(m\) denotes the neutron mass, \(\lambda_{\mathrm{d}}=2\pi\hbar/(mv)\) is its de Broglie wavelength, \(g\) is the local gravitational acceleration, \(l\) is the vertical height of the parallelogram, and \(s\) is the length of the baseline AB.

%\begin{figure}[h]
%		\hspace*{-1.3cm}
    %\centering
%	\includegraphics[scale=0.5]{COW.pdf}%0.13
%	\caption{Schematic of the COW experiment: a neutron beam is coherently split along the two sides of a vertically oriented parallelogram and subsequently recombined to produce interference.}
%	\label{COW experiment}
%\end{figure}

\begin{figure}[h]
	\hspace*{-1.3cm}
    %\centering
	\includegraphics[scale=0.5]{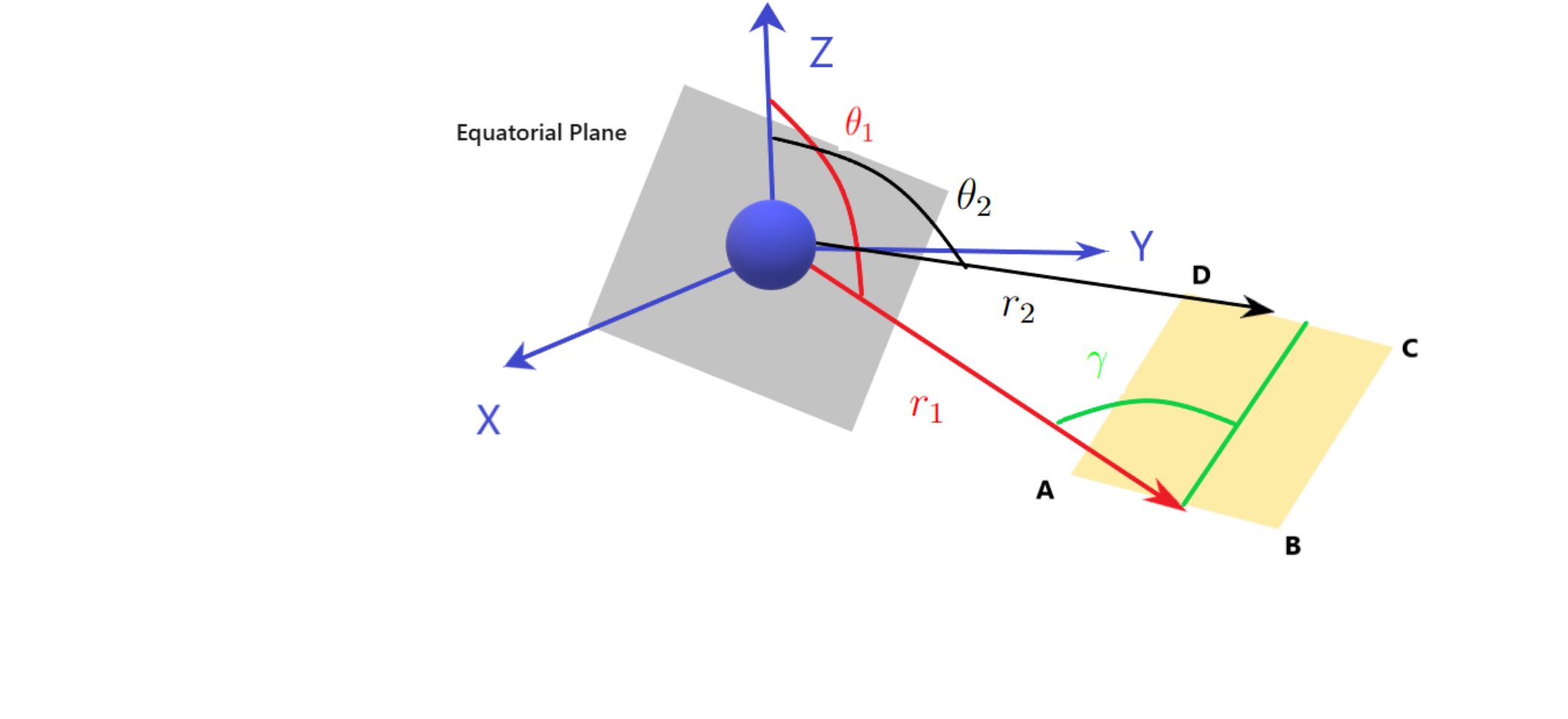}%0.13
	\caption{ The interference experiment is conducted in the region of $r \gg r_g$ within the EMS spacetime. Particles are split into two beams traversing paths ABC and ADC, respectively, before interfering with each other. The X and Y axes lie on the equatorial plane, which is the accretion disk.
   % The $z$-axis of the black hole, together with vectors $\vec{r}_1$ and $\vec{r}_2$, lies in a common plane that is perpendicular to the base AB. 
    The angle between $\vec{r}_1$ and the plane of the parallelogram ABCD is denoted as $\gamma$. The length of segment AB is $s$, and the height of the parallelogram is $l$.   }
	\label{experimentEMS}
\end{figure}

We next embed the parallelogram interferometer in the asymptotically flat region (\(r\gg r_{g}\)) of an EMS black-hole spacetime, allowing for arbitrary massive probe particles; the corresponding geometry is shown in Fig.~\ref{experimentEMS}. To render the calculation analytically tractable we adopt two idealised conditions:

(a)Local homogeneity-—the linear extent of the interferometer is much smaller than its coordinate distance from the horizon, so that the metric functions \(r\) and \(\theta\) remain effectively constant along the segments AB and DC;

(b)Reflection-time energy conservation-—the Killing energy \(\mathcal{E}\) is unchanged when the particle reverses direction at vertices B and D, whereas the angular momentum \(\mathcal{L}\) is instantaneously readjusted at these vertices yet conserved separately along each straight segment (AB, BC, AD, DC); consequently the magnitude of the local velocity, defined by Eq.~\eqref{vsquare}, is unaltered by the reflections.

Combining the assumption (a) with \eqref{PhiK}, we write the accumulated 
gravitational phase along the path AB as 
\bea
\phi_{g,AB} &\approx&
\frac{1}{\hbar}\Bigl(
\int S_0 \dd t +\int S_3 \dd \phi 
\Bigr)
\nonumber\\
&=&\frac{1}{\hbar}
(S_0 t_{AB} + S_3 \phi_{AB})
\nonumber\\
&=&
\frac{1}{\hbar} \mathcal{E} (1-\gamma_{00}^{-1}) t_{AB}
+\frac{1}{\hbar} \Bigl[
\mathcal{L}_{\rm AB}
(1-\gamma_{22}^{-1}r) 
\Bigr] \phi_{AB},
\label{phiAB0}
\eea
where $t_{AB}=t_B-t_A$ and $\phi_{AB}=\phi_B-\phi_A$ are defined.
We only consider the case $M \ll r$,
therefore, expanding \eqref{phiAB0} at the third order, we get:
\be
\phi_{g,AB} \approx
\frac{1}{\hbar} \mathcal{E} 
t_{AB}\Bigl[-\frac{M}{r_1}+\frac{Q^2(1-\alpha^2+\beta)}{2r_1{^2}}\Bigr]
+\frac{1}{\hbar} \phi_{AB}\mathcal{L}_{\rm AB}
\Bigl[-\frac{\alpha^2 Q^2}{2M r_1}-\frac{3}{2}\Bigl(\frac{\alpha^2 Q^2}{2M 
r_1}\Bigr)^{2}
\Bigr].
\label{phiAB}
\ee
The quantity $\mathcal{L}$ is given by (see Appendix~\ref{time and angle})
\be
\mathcal{L}=\frac{\mathcal{E}v^\phi \Gamma_{33}  }{\sqrt{g_{00}}},
\label{L given}
\ee
where $v^j$ is the three-dimensional velocity and $\Gamma_{ij}$ is the 
three-dimensional metric tensor defined by~\cite{Landau:1975pou}
\be
v^k=\frac{\dd x^k}{\sqrt{g_{00}} (\dd x^0+\frac{g_{0i}}{g_{00}}
	\dd x^i)},\qquad
\Gamma_{ij} =-g_{ij}+\frac{g_{0i}g_{0j}}{g_{00}}.
\label{v components}
\ee
For the gravitational phases $\phi_{g,BC}$, $\phi_{g,AD}$, and $\phi_{g,DC}$, 
the derivation is similar to $\phi_{AB}$. Combining these phases, we can get 
the phase difference between the paths ADC and ABC. Hence, with the following 
relations (see Appendix~\ref{time and angle}):
\be
r_2\approx r_1 +\frac{l\cos(\gamma)}{\sqrt{-g_{11}(r_1,\theta_1)} },
\qquad
\theta_2\approx \Bigl| \theta_1-
\frac{l\sin(\gamma)}{\sqrt{-g_{22}(r_1,\theta_1)}} \Bigr|,
\label{r2 theta2}
\ee
we can expand the phase difference in the neighborhoods of $r_1$ and 
$\theta_1$, and for simplicity we only keep the first order terms of $l/r_1$. 
Then relate the time, the angle, and the energy with the observations 
(Appendix~\ref{time and angle}):
\bea
&&t_{AB}\approx 
\frac{s}{v\sqrt{g_{00} }},
\quad
\phi_{AB}\approx\frac{s}{\sqrt{\Gamma_{33} }},
\label{t varphi}
\\
&&\mathcal{E}=
\begin{cases}
	m(1-v^2)^{-1/2} \sqrt{g_{00}}, &\text{for massive particles},\\
	\hbar \omega \sqrt{g_{00}}, &\text{for massless particles}.
\end{cases}
\label{mathcalE lambda}
\eea
With the above steps, we derive
the phase difference between the paths ADC and ABC as follows (see 
Appendix~\ref{DPD} for more details)
\begin{equation}
\begin{aligned}
\delta\phi_{g} \approx\ &
\frac{\mathcal{E}_{0} l s}{\hbar r_1} \Biggl\{
\frac{1}{v} \left[
\frac{M \cos\gamma}{r_1} 
- \frac{\cos\gamma Q^2 (1 - \alpha^2 + \beta)}{r_1^2}
\right]  \\
&\quad + \frac{v}{l} \Biggl[
\frac{l Q^2 \alpha^2}{2 M r_{1}} \cos\theta_1 \sin\gamma 
+ \frac{1}{8 M r_{1}^2} \bigg(
3 l Q^4 \alpha^4 \cos\theta_1 \sin\gamma \\
&\qquad\qquad
+ (4 l M Q^2 \alpha^2 + 3 l Q^4 \alpha^4 
+ 4 l^2 Q^2 \alpha^2 \cos\gamma) \cos\gamma \sin\theta_1
\bigg)
\Biggr]
\Biggr\},
\end{aligned}
\label{PhaseDifference total}
\end{equation}
and $\mathcal{E}_0$ is defined by:
\be
\mathcal{E}_0=
\begin{cases}
	m(1-v^2)^{-1/2}, &\text{for massive particles},\\
	\hbar \omega, &\text{for massless particles}.
\end{cases}
\ee

In particular, for $\gamma=0$ and $\gamma=\pi/2$, the gravitational phase 
differences are respectively:
\begin{equation}
	\begin{aligned}
		\delta\phi_{g}| _{\gamma=0}
		\approx\ 
		\frac{\mathcal{E}_{0} l s}{\hbar r_1} \Biggl\{
		\frac{1}{v} \left[
		\frac{M}{r_1} 
		- \frac{Q^2 (1 - \alpha^2 + \beta)}{r_1^2}
		\right] 
		+ \frac{v}{l} \cdot \frac{1}{8 M r_{1}^2} \left(
		4 l M Q^2 \alpha^2 + 3 l Q^4 \alpha^4 + 4 l^2 Q^2 \alpha^2
		\right) \sin\theta_1
		\Biggr\},
	\end{aligned}
\end{equation}

\begin{equation}
	\begin{aligned}
		\delta\phi_{g}|_{\gamma=\frac{\pi}{2}}
		\approx\ 
		& \frac{\mathcal{E}_{0} l s}{\hbar r_1} \Biggl\{
		\frac{v}{l} \Biggl[
		\frac{l Q^2 \alpha^2}{2 M r_{1}} \cos\theta_1
		+ \frac{1}{8 M r_{1}^2} \cdot 3 l Q^4 \alpha^4 \cos\theta_1
		\Biggr]
		\Biggr\}.
	\end{aligned}
\end{equation}

The prediction \eqref{PhaseDifference total} can be tested experimentally by 
measuring the fringe shift as a function of $\gamma$. 
As for the non-relativistic particles, \eqref{PhaseDifference total} is reduced 
to
\begin{equation}
\begin{aligned}
\delta\phi_{g}^{\rm NR} \approx\ 
& \frac{m l s}{\hbar r_1} \Biggl\{
\Bigl(\frac{1}{v}+\frac{v}{2}\Bigr) \left[
\frac{M \cos\gamma}{r_1} 
- \frac{\cos\gamma Q^2 (1 - \alpha^2 + \beta)}{r_1^2}
\right] \\
& \quad + \frac{v}{l} \Biggl[
\frac{l Q^2 \alpha^2}{2 M r_{1}} \cos\theta_1 \sin\gamma 
+ \frac{1}{8 M r_{1}^2} \bigg(
3 l Q^4 \alpha^4 \cos\theta_1 \sin\gamma \\
& \quad\quad + (4 l M Q^2 \alpha^2 + 3 l Q^4 \alpha^4 + 4 l^2 Q^2 \alpha^2 
\cos\gamma) 
\cos\gamma \sin\theta_1
\bigg)
\Biggr]
\Biggr\},
\end{aligned}
\label{PhaseDifference NR}
\end{equation}
where we have neglected the terms of $O(v^2)$ and higher orders.

%%

%%%%%
\subsection{Numerical simulations and discussion}
For simplicity,   we restore the SI units. 
From \eqref{PhaseDifference total}, we can find 
$\delta\phi(\gamma=\pi)=-\delta\phi(\gamma=0)$. Therefore, if we change the 
angle $\gamma$ from $0$ to $\pi$, we get a fringe shift:
\be
N=\Bigl| \frac{\delta\phi_{g}(\gamma=\pi)-\delta\phi_{g}(\gamma=0)}{2\pi}\Bigr|
=\Bigl| \frac{\delta\phi_{g}(\gamma=0)}{\pi}\Bigr|.
\label{defineN}
\ee
For the setup of experiment, we take the parameters in \cite{Overhauser:1974}, 
i.e.
\be
l=2\times 10^{-2} {\rm m}^2,
\qquad
s=3\times 10^{-2} {\rm m}^2,
\qquad
\lambda_d=1.42\times 10^{-10} {\rm m}.
\label{lambda d}
\ee
And for the constants in the Eq.(\ref{PhaseDifference total}), we list them  
below\cite{Carroll}: 
\begin{equation}
	\hbar\approx1.055\times10^{-34}J\cdot s,
	\qquad
	c=3.0\times 10^{8} m/s
	\qquad
	G=6.67\times 10^{-11} m^{3}\cdot kg^{-1}\cdot s^{-2},
	\qquad
	\epsilon_0=8.85\times 10^{-12}\rm F/m.
\end{equation}

\subsubsection{Black hole with no charge}
In this part, we will consider the black hole with no charge.  From the 
Eq.(\ref{PhaseDifference total}), we can easily know that the gravitational 
phase difference is totally dominated by the first term.  In the following we 
discuss two examples in which the particles that interfere are neutrons. 

(I)At first, we will take the earth as the gravitational source.  For the 
parameter $r_1$, we assume the equatorial radius of the earth. 
Therefore, we get the results:
\be
\delta\phi_g=33.5498.
\label{mass I}
\ee
As for the fringe according to \eqref{defineN}, we get:
\be
N=10.6792. \label{earthN}
\ee
The value \eqref{earthN} is nearly the same as the result 
in~\cite{Overhauser:1974,203}, which agrees with the claim that the equation 
\eqref{PhaseDifference total} produces the result \eqref{COWpre} on the earth 
in the Newtonian limit.

(II) We take $\rm M87^{*}$ as the source of black hole. And we take the 
distance between the black hole and the earth to be the value of $r_1$. The 
parameters are given by~\cite{204} 
\be
M=6.5\times 10^{9} M_\odot,
\qquad
r_1=16.8{\rm Mpc},
\label{parameters X1}
\ee
where $M_\odot$ is the mass of the sun.  Thus we get:
\be
\delta\phi_g=9.803\times 10^{-18}.
\label{mass II}
\ee
For the fringe shift we get:
\be
N=3.12\times 10^{-18},
\label{nmass number}
\ee
corresponding to changing the angle $\gamma$ from $0$ to $\pi$, which is much 
smaller than the fringe shift in the example (I). 

\subsubsection{Black hole with charge}
In this part, we will consider the black hole with charge. Similar to above, 
this part we only  discuss two examples which will tell us charge how affect 
the gravitational phase. 

%In paper\cite{205}, they calculate the upper limit of charge $(\approx 10^{15} 
%C)$ for $Sgr {\rm A^*}$. 
(I)For simplicity, we here assume $\theta_{1}=0$, then we have:
\begin{equation}
	\delta\phi_{g}^{\mathrm{NR}}
	\approx
	\frac{m l s}{\hbar r_1}\left(\frac{1}{v}+\frac{v}{2}\right)
	\left(\frac{\rm M}{r_1}-\frac{Q^2(1-\alpha^2+\beta)}{r_1^2}\right).
\end{equation}
Now, we consider the concrete form in SI units:
\begin{equation}
	\delta \phi_g^{\mathrm{NR}} \approx \frac{m l s c}{\hbar 
	r_1}\left(\frac{mc}{v}+\frac{v}{2mc}\right)\left(\frac{G M}{c^2 
	r_1}-\frac{G Q^2\left(1-\alpha^2+\beta\right)}{4 \pi \epsilon_0 c^4 
	r_1^2}\right).
\end{equation}
We take the $\rm Sgr A^{*}$ as the source of black hole, other parameters are given by\cite{206},
\begin{equation}
	\rm M=4.297\times 10^{6} M_\odot,
	\qquad
	r_1= 8.275 \rm  kpc.
\end{equation}
Then, we get the results:
\begin{equation}
	\delta \phi= 3.0\times 10^{-14}-1.4\times 10^{-78}\times Q^{2}\times 
	(1-\alpha^2+\beta).
\end{equation}
And the fringe shift is :
\begin{equation}
	N=9.55\times10^{-15}-4.38\times10^{-79}\times Q^2 \times (1-\alpha^2+\beta).
\end{equation}
It seems like charge has no contributions to the gravitational phase.

(II)However, if we consider $\gamma=\pi/4,\theta_{1}=0$, thus we have:
\begin{equation}
	\delta \phi_g^{\mathrm{NR}} \approx \frac{m l s}{\hbar r_1} 
	\frac{\sqrt{2}}{2}\left\{\left(\frac{1}{v}+\frac{v}{2}\right)\left[\frac{M}{r_1}-\frac{Q^2\left(1-\alpha^2+\beta\right)}{r_1^2}\right]+v\left[\frac{Q^2
	 \alpha^2}{2 M r_1}+\frac{3 Q^4 \alpha^4}{8 M r_1^2}\right]\right\}.
\end{equation}
Now we consider extremally charged black holes, 
\begin{equation}
	Q_{\mathrm{ext}}=M \sqrt{4 \pi \epsilon_0 G}=7.36\times 10^{26} \rm C.
\end{equation}
Thus we have:
\begin{equation}
	\delta \phi_g^{\mathrm{NR}} \approx 2.11 \times 10^{-14} + 2.89 \times 
	10^{-69} \, Q_{\mathrm{ext}}^2 \alpha^2\approx 2.11\times 
	10^{-14}+2.89\times 10^{-17}\alpha^2.
\end{equation}
And we can have the fringe shift:
\begin{equation}
	\tilde{N}=\Big|4.5\times 10^{-18}\alpha^2-1.4\times 10^{-15}\Big|,
\end{equation}
corresponding to changing the angle $\gamma$ from $0$ to $\pi/4$.

\subsubsection{Discussion}
The numerical results for phase differences under different gravitational 
sources clearly demonstrate the dominant role of the mass and the 
source-detector distance in determining the interference phase shift. In the 
case of an uncharged black hole, taking the Earth as the gravitational source 
yields a phase difference of approximately 33.55 and a fringe shift 
$N \approx 10.68$, consistent with classical neutron interferometry experiments 
such as those by Overhauser et al., thereby validating the theoretical 
formulation and its experimental feasibility. By contrast, for supermassive 
black holes such as $M87^{*}$, despite their enormous mass ($6.5 \times 10^{9} 
M_{\odot}$), the extremely large distance (16.8 Mpc) reduces the phase 
difference to the order of $10^{-18}$ and fringe shift to $3.12 \times 
10^{-18}$, effectively making experimental detection infeasible with current 
technology.

When considering charged black holes, although the direct contribution of the 
black hole charge $Q$ to the gravitational phase difference is typically 
several orders of magnitude smaller than the dominant mass term—for example, in 
the case of Sgr A\* ($M=4.3 \times 10^{6} M_{\odot}$, 
$r_1=8.275\,\mathrm{kpc}$), the charge correction term is on the order of 
$10^{-78} \times Q^2$ and thus negligible—the analysis reveals that this 
contribution is modulated by the coupling parameter $\alpha$. Specifically, for 
extremal charges $Q_{\mathrm{ext}} \approx 7.36 \times 10^{26}\, \mathrm{C}$, 
the charge-related phase shift correction can reach up to $2.89 \times 10^{-17} 
\alpha^{2}$, which, while small, is many orders of magnitude larger than the 
charge contributions without $\alpha$-modulation.

This implies that if the coupling parameter $\alpha$ attains values of order 
unity or larger, the charge-induced effect on the gravitational phase 
difference could be enhanced sufficiently to be within reach of highly 
sensitive quantum interference measurements. Consequently, tuning or 
constraining $\alpha$ presents a promising avenue for indirect experimental 
detection of black hole charge effects, which are otherwise obscured by the 
overwhelming dominance of mass contributions.

\section{Conclusion\label{Conclusion}}
In this work, we have formulated the gravitational phase difference as the 
integral of a function $S_\beta$, defined as the product of the four-momentum, 
the metric tensor, and the gauge gravitational potential—where the latter is 
naturally expressed within the tetrad formalism. This formulation provides a 
general and covariant framework for evaluating gravitational phases in 
arbitrary curved spacetimes.

As a concrete application, we have examined the Einstein–Maxwell–Scalar (EMS) 
spacetime, analyzing a particle interferometry setup (FIG.~\ref{experimentEMS}) 
analogous to the classic Colella–Overhauser–Werner (COW) experiment, but 
adapted to the EMS black hole background. Within this framework, we have 
computed and compared the phase differences for black holes with and without 
electric charge.

Our results confirm that, among the three fundamental black hole 
parameters—mass, angular momentum, and electric charge—the dominant 
contribution to the gravitational phase difference arises from the black hole 
mass. Nevertheless, in the case of certain extremal black holes and for 
specific parameter regimes, the electric charge can produce a comparable or 
even dominant effect. Notably, the coupling parameter $\alpha$ in the EMS model 
plays a crucial role in modulating these effects, potentially amplifying 
otherwise negligible charge-induced contributions.

From a broader perspective, while the influence of black hole charge on quantum 
phases is typically small in astrophysical scenarios, our analysis demonstrates 
that non-minimal couplings can enhance these signatures to potentially 
detectable levels. This opens a new theoretical window for probing black hole 
charge and related couplings via high-precision quantum interferometry. Future 
work may extend this approach to rotating black holes, dynamical spacetimes, 
and alternative theories of gravity, offering a pathway toward experimentally 
testing subtle quantum–gravitational interactions in strong-field regimes.

\acknowledgments
We are especially grateful to Qiang Li for many stimulating and helpful discussions on this project. During the preparation of this manuscript, the authors utilized AI tools (e.g., Kimi) to assist in polishing and rephrasing the language of certain sections. All academic content, modeling ideas, and cited sources have been clearly acknowledged. The AI was not involved in research design, data interpretation, or the generation of novel viewpoints. The authors take full responsibility for the final content of the paper.
%This work was supported by the National Key Research and Development Program of China (No. 2022YFA1403700), NSFC (Grants No. 12334002), Guangdong Provincial Quantum Science Strategic Initiative Grand No. SZZX2401001, the Science, Technology and Innovation Commission of Shenzhen Municipality (No. ZDSYS20190902092905285), and Center for Computational Science and Engineering at Southern University of Science and Technology. 
This work was supported by the National Key Research and Development Program of China (No. 2024YFA1408101), NSFC (Grants No. 12334002), Guangdong Provincial Quantum Science Strategic Initiative Grant No. SZZX2401001, the SUSTech-NUS Joint Research Program, the Science, Technology and Innovation Commission of Shenzhen Municipality (No. ZDSYS20190902092905285), and Center for Computational Science and Engineering at Southern University of Science and Technology.

%%%%%%%%
\appendix 
\section{Derivations for some formulas\label{time and angle}}
In this appendix we derive some equations used in Sec.~\ref{interference}. Let 
us prove \eqref{L given} firstly. According to \eqref{conserved quantities}, we 
get
\be
\mathcal{L}=-\mathcal{E}\frac{g_{33}}{g_{00}}\frac{\dd\phi}{\dd t}.
\label{L matric}
\ee
The expression for $\dd \phi/\dd t$ is found by letting $k=3$ in the definition 
of the velocity \eqref{v components}, namely 
\be
\frac{\dd \phi}{\dd t}
=v^\phi \sqrt{g_{00}} .
\label{dtdphi}
\ee 
Hence, replacing \eqref{dtdphi} into \eqref{L matric}, we prove \eqref{L given}.

We now derive Eq.~\eqref{r2 theta2}. Following Ref.~\cite{Landau:1975pou,203}, in any stationary spacetime the spatial distance \(L\) between two neighbouring points is obtained from the three-dimensional line element:
\be
\dd L^2 =\Gamma_{ij}\dd x^i \dd x^j,
\label{dLsquare}
\ee
where the three-dimensional metric tensor $\Gamma_{ij}$ is defined in \eqref{v 
components}.  Projecting this element along the radial direction and perpendicular to it (cf.~Fig.~\ref{experimentEMS}) yields
\bea
&&l\cos(\gamma) =\sqrt{\Gamma_{11}} \dd r
\approx \sqrt{-g_{11}} (r_2-r_1),
\\
&&l \sin(\gamma) =\bigl |\sqrt{\Gamma_{22}} \dd \theta \bigr|
\approx
\begin{cases}
	\sqrt{-g_{22}} (\theta_1-\theta_2),\quad &\text{when 
	$l\sin(\gamma)\sqrt{-g_{22}^{-1}}\le  \theta_1$,}\\
	\sqrt{-g_{22}} (\theta_1+\theta_2)
	\quad &\text{when $l\sin(\gamma)\sqrt{-g_{22}^{-1}}>  \theta_1$,}
\end{cases}
\eea
respectively, from which the two relations in Eq.~\eqref{r2 theta2} follow immediately.

To obtain Eq.~\eqref{t varphi}, we note that assumption (a) of Sec.~\ref{TPre} implies $\dd r\approx 0$, $\dd\theta\approx 0$ along AB.  Consequently, Eq.~\eqref{dLsquare} reduces to
\be
\dd\phi\simeq\frac{\dd L}{\sqrt{\Gamma_{33}}}.
\label{dvarphi 1}
\ee
Combining this relation with Eq.~\eqref{dtdphi} yields the link between coordinate time and infinitesimal proper length,
\be
\dd t\simeq\frac{\dd L}{v^{\phi}\sqrt{g_{00}\Gamma_{33}}}.
\label{ds 1}
\ee
Integrating the two expressions and imposing the boundary condition \eqref{v varphi 0} immediately supplies the pair of formulae collected in Eq.~\eqref{t varphi}.

For the energy of a massive particle in \eqref{mathcalE lambda}, the result is directly adopted from Ref.~\cite{Landau:1975pou,203} (see the corresponding derivation in~\cite{Landau:1975pou}). In contrast, the energy of a massless particle is given by~\cite{Carroll:2004st}:  
\be  
\mathcal{E} = V \hbar \omega,  
\label{EV}  
\ee  
where $\omega$ denotes the frequency measured by a static observer, and $V$ represents the redshift factor expressed as:  
\be  
V = \sqrt{K_\mu K^\mu},  
\label{redshift factor}  
\ee  
with $K^\mu$ being the Killing vector associated with time-translation invariance. A static observer here is defined such that the four-velocity is proportional to the Killing vector~\cite{Carroll:2004st}. Substituting the Killing vector $K^\mu = (1, 0, 0, 0)$ into \eqref{redshift factor}, we find $V = \sqrt{g_{00}}$. Inserting this result into \eqref{EV} yields the second expression in \eqref{mathcalE lambda}.

%%%%%%

%%%%%
\section{Derivations for the phase difference\label{DPD}}
In this appendix, we detail the derivation of the phase difference \eqref{PhaseDifference total} between the paths ADC and ABC in FIG.~\ref{experimentEMS}. Begin by specifying the coordinates of points A, B, C, and D as:  
\be  
{\rm A}(t_A, r_1, \theta_1, \phi_A), \quad  
{\rm B}(t_A+t_{AB}, r_1, \theta_1, \phi_A+\phi_{AB}), \quad  
{\rm C}(t_D+t_{DC}, r_2, \theta_2, \phi_D+\phi_{DC}), \quad  
{\rm D}(t_D, r_2, \theta_2, \phi_D),  
\label{coordinatesABCD}  
\ee  
where $t_{AB} \equiv t_B - t_A$, $\phi_{AB} \equiv \phi_B - \phi_A$, $t_{DC} \equiv t_C - t_D$, and $\phi_{DC} \equiv \phi_C - \phi_D$ are defined for notational clarity.  

Consequently, the phases for each path are formulated as follows. Under assumption (a) stated in Sec.~\ref{TPre}, the conditions $\dd r \approx 0$ and $\dd\theta \approx 0$ hold for paths AB and DC. Thus, Eq.~\eqref{PhiK} reduces to:  
\bea  
\phi_{g,AB} &\approx& \frac{1}{\hbar} \left( S_0^A t_{AB} + S_3^A \phi_{AB} \right)_{AB},  
\label{phaseAB} \\  
\phi_{g,DC} &\approx& \frac{1}{\hbar} \left( S_0^D t_{DC} + S_3^D \phi_{DC} \right)_{DC},  
\label{phaseDC}  
\eea  
where superscripts A and D signify the respective positions, and subscripts AB and DC denote the paths under consideration.  

For path AD, the phase is expressed as:  
\begin{align}  
\phi_{g,AD} 
&= \frac{1}{\hbar}\Bigl(\int S_\beta \dd x^\beta \Bigr)_{AD} \nonumber \\  
&= \frac{1}{\hbar}\Bigl[
S_0 (\vec{r}_a) t_{AD} + S_1 (\vec{r}_b) r_{AD}+ S_2 (\vec{r}_c) \theta_{AD} + 
S_3(\vec{r}_d) \phi_{AD}
\Bigr]_{AD} \nonumber \\  
&= \frac{1}{\hbar}\Bigl\{
\bigl[ S_0 (\vec{r}_a) - S_0^A + S_0^A \bigr] t_{AD}
+ \bigl[ S_1 (\vec{r}_b) - S_1^A + S_1^A \bigr] r_{AD} \nonumber \\  
&\quad + \bigl[ S_2 (\vec{r}_c) - S_2^A + S_2^A \bigr] \theta_{AD}
+ \bigl[ S_3(\vec{r}_d) - S_3^A + S_3^A \bigr] \phi_{AD}
\Bigr\}_{AD} \nonumber \\  
&\approx \frac{1}{\hbar} \bigl[ S_0^A (t_D-t_A) + S_1^A (r_2-r_1) + S_2^A 
(\theta_2-\theta_1) + S_3^A (\phi_D-\phi_A) \bigr]_{AD},  
\label{phaseAD}  
\end{align}  
where the mean value theorem for integrals is employed in the second step, and $\vec{r}_a$, $\vec{r}_b$, $\vec{r}_c$, and $\vec{r}_d$ denote points along AD. The final step in \eqref{phaseAD} is valid because both $(S_\beta(\vec{r}_p)-S_\beta^A)_{AD}$ and $x^\beta_{AD}$ are bounded by $O(l/r_1^2)$. Their expressions are at least proportional to $1/r$, and upon Taylor expansion, their differences are of order $1/r^2$ or higher.

Following a similar derivation to that for \eqref{phaseAD}, we obtain:  
\bea  
\phi_{g,BC}&\approx& \frac{1}{\hbar} \bigl[ S_0^B (t_D-t_A +t_{DC}-t_{AB})
+S_1^B (r_2-r_1) +S_2^B (\theta_2-\theta_1) +S_3^B (\phi_D-\phi_A 
+\phi_{DC}-\phi_{AB} ) 
\bigr]_{BC},  
\label{phaseBC}  
\eea  
with \eqref{coordinatesABCD} invoked in the derivation. To simplify the expression for $(\phi_{g,AD}-\phi_{g,BC})$, we first establish the equality $(S_\beta^A)_{AD}=(S_\beta^B)_{BC}$. Notably, $(g_{\mu\nu}^A)_{AD}=(g_{\mu\nu}^B)_{BC}$ holds due to the time- and $\varphi$-independence of the metric. Incorporating this result and the expression for $S_\beta$ in \eqref{S beta2}, we need to verify the following equalities:  
\be  
(P^1)^A_{AD}=(P^1)^B_{BC}, \qquad  
(P^2)^A_{AD}=(P^2)^B_{BC}, \qquad  
(\mathcal{E}^A)_{AD}=(\mathcal{E}^B)_{BC}, \qquad  
(\mathcal{L}^A)_{AD}=(\mathcal{L}^B)_{BC}.  
\label{proof}  
\ee  

The equalities in \eqref{proof} hold due to assumptions (a) and (b) in Sec.~\ref{TPre}, which further imply $(\vec{v}_A)_{AD}=(\vec{v}_B)_{BC}$ for particles on the parallelogram. Combining this with \eqref{L given} confirms the last equality in \eqref{proof}. With \eqref{proof} verified, the relation $(S_\beta^A)_{AD}=(S_\beta^B)_{BC}$ is established. Together with \eqref{phaseAD} and \eqref{phaseBC}, this yields:  
\be  
\phi_{g,AD}-\phi_{g,BC}=-\frac{1}{\hbar}\bigl[
S_0^B (t_{DC}-t_{AB})
+S_3^B (\phi_{DC}-\phi_{AB} )
\bigr]_{BC}.  
\label{ADBC}  
\ee  

Combining \eqref{ADBC}, \eqref{phaseAB}, and \eqref{phaseDC}, the total phase difference between paths ADC and ABC is:  
\bea  
\delta\phi_{g}
&=&\phi_{g,ADC}-\phi_{g,ABC}
\nonumber\\
&=&\phi_{g,DC}-\phi_{g,AB}+\phi_{g,AD}-\phi_{g,BC}
\nonumber\\
&=& \frac{1}{\hbar}\biggl\{
\bigl[ (S_0^D)_{DC} -(S_0^A)_{AB} \bigr] t_{AB} 
+\bigl[ (S_3^D)_{DC} -(S_3^A)_{AB} \bigr] \phi_{AB}
\nonumber\\
&&
+\bigl[(S_0^D)_{DC} -(S_0^B)_{BC} \bigr] (t_{DC}-t_{AB})
+\bigl[(S_3^D)_{DC} -(S_3^B)_{BC} \bigr] (\phi_{DC}-\phi_{AB})
\biggr\}.  
\label{delta phi0}  
\eea  

From the expression of $S_\beta$ in \eqref{S beta2}, $S_0$ is independent of $\mathcal{L}$, implying that $S_0$ remains invariant when the probe particle reverses direction at points B and D. Thus, $(S_0^B)_{BC} = (S_0^B)_{AB} = (S_0^A)_{AB}$. Incorporating \eqref{phaseAB}, \eqref{delta phi0} can be rewritten as:  
\be  
\delta\phi_g = \delta\phi_{g,1} + \delta\phi_{g,2} + \delta\phi_{g,3},  
\label{delta phi01}  
\ee  
where  
\bea  
\delta\phi_{g,1} &=& \frac{1}{\hbar}\bigl[ (S_0^D)_{DC} t_{AB} + (S_3^D)_{DC}\phi_{AB}\bigr] - \phi_{g,AB},  
\label{delta phi a} \\  
\delta\phi_{g,2} &=& \frac{1}{\hbar}\bigl[(S_0^D)_{DC} - (S_0^A)_{AB} \bigr] (t_{DC}-t_{AB}),  
\label{delta phi b} \\  
\delta\phi_{g,3} &=& \frac{1}{\hbar}\bigl[(S_3^D)_{DC} - (S_3^B)_{BC} \bigr] (\phi_{DC}-\phi_{AB}).  
\label{delta phi c}  
\eea

We now explicitly evaluate these phase differences. Note that $\phi_{g,AB}$ is given in \eqref{phiAB}. For the first term in \eqref{delta phi a}, comparison with \eqref{phaseAB} shows that it can be obtained by substituting $r_1 \to r_2$, $\theta_1 \to \theta_2$, and $\mathcal{L}_{AB} \to \mathcal{L}_{DC}$ in \eqref{phiAB}. Thus, \eqref{delta phi a} is equivalent to:  
\be  
\delta\phi_{g,1} = \phi_{g,AB}(r_1 \to r_2, \theta_1 \to \theta_2, \mathcal{L}_{AB} \to \mathcal{L}_{DC}) - \phi_{g,AB}(r_1, \theta_1, \mathcal{L}_{AB}).  
\label{delta a1}  
\ee

Next, we derive $\mathcal{L}_{AB}$. To this end, we first determine $v^\phi$ using \eqref{L given}. From the relation in \cite{Landau:1975pou,203}:  
\be  
v^2 = \Gamma_{ij} v^i v^j,  
\label{vsquare}  
\ee  
in the EMS spacetime, this gives:  
\be  
|v^\phi| = \sqrt{\frac{v^2 + g_{11} (v^r)^2 + g_{22} (v^\theta)^2}{\Gamma_{33}}}.  
\label{v varphi}  
\ee  

For path AB, with $v^\phi > 0$, $v^r = 0$, and $v^\theta = 0$, \eqref{v varphi} simplifies to:  
\be  
v^\phi = \sqrt{\frac{v^2}{\Gamma_{33}}}.  
\label{v varphi 0}  
\ee  

Substituting \eqref{v varphi 0} into \eqref{L given} yields:  
\be  
\mathcal{L}_{AB} = \frac{\mathcal{E}}{g_{00}}\Bigl( \sqrt{g_{00} \Gamma_{33} v^2} \Bigr) \approx \mathcal{E} r_1 \sin(\theta_1) \epsilon,  
\label{LAB}  
\ee  
where $\epsilon$ is defined as:  
\be  
\epsilon = \begin{cases} 
\sqrt{1 - \frac{m^2}{\mathcal{E}^2}}, & \text{for massive particles}, \\ 
1, & \text{for massless particles}. 
\end{cases}  
\label{epsilon}  
\ee  

For $\mathcal{L}_{DC}$, one simply replaces $r_1$ with $r_2$ and $\theta_1$ with $\theta_2$ in \eqref{LAB}.  

As for \eqref{delta a1}, using \eqref{r2 theta2} we expand $\delta\phi_{g,1}$ 
in the neighborhoods of $r_1$ and $\theta_1$ up to the first order. Finally, we 
get:
\bea
\delta\phi_{g,1}
&\approx&
\frac{\mathcal{E}_{0} ls}{\hbar r_1} \Bigl\{
\frac{1}{v}\Bigl[\frac{M \cos\gamma}{r_1}-\frac{\cos\gamma 
Q^2(1-\alpha^2+\beta)}{r^{2}_{1}}
\Bigr]+
\frac{v\Delta}{l } 
\Bigr\},
\label{PhaseDifference a}
\eea
where $\Delta$ is defined by:
\begin{equation}
	\begin{aligned}
		\Delta \approx\ 
		\frac{l Q^2 \alpha^2}{2 M r_{1}} \cos\theta_1 \sin\gamma 
		+ \frac{1}{8 M r_{1}^2} \bigg[
		3 l Q^4 \alpha^4 \cos\theta_1 \sin\gamma
		+ (4 l M Q^2 \alpha^2 + 3 l Q^4 \alpha^4 + 4 l^2 Q^2 \alpha^2 
		\cos\gamma) \cos\gamma \sin\theta_1
		\bigg],
	\end{aligned}
\end{equation}
and $\mathcal{E}_0$ is defined by:
\be
\mathcal{E}_0=
\begin{cases}
	m(1-v^2)^{-1/2}, &\text{for massive particles},\\
	\hbar \omega, &\text{for massless particles}.
\end{cases}
\label{mathcalE lambda00}
\ee

As for $\delta\phi_{g,2}$ and  $\delta\phi_{g,3}$ in \eqref{delta phi b} and 
\eqref{delta phi c}, they are at least $O(l^3/r_1^3)$ and higher orders
because both $[(S_0^D)_{DC} -(S_0^A)_{AB}]$ and $(t_{DC}-t_{AB})$ are of the 
order of $O(l^2/r_1^2)$ and higher orders, according to \eqref{r2 theta2} and 
\eqref{t varphi}. Therefore, in our approximation (we remind the reader the 
sentence after \eqref{r2 theta2}), this phase difference is negligible:
\be
\delta\phi_{g,2}\approx 0,
\label{PhaseDifference cc}
\quad
\delta\phi_{g,3} \approx 0.
\ee

%%%%%%%%%%

\bibliographystyle{unsrt}
\bibliography{TG}

\end{document}